\documentclass[a4paper]{article}
\usepackage{Odyssey2022}
\usepackage{epsfig,amssymb,amsmath}
\usepackage{multirow,array,enumitem,adjustbox}
\usepackage{booktabs, threeparttable,caption, makecell}
\usepackage[hyphens]{url}
\usepackage{hyperref}
\hypersetup{
breaklinks=true,   
colorlinks=true,   
linkcolor=black,
anchorcolor=black,
citecolor=black,
filecolor=black,
menucolor=black,
runcolor=black,,
urlcolor=red,
        }
\usepackage{cleveref}

\ninept

\title{The SJTU X-LANCE Lab System for CNSRC 2022}

\name{Zhengyang Chen, Bei Liu, Bing Han, Leying Zhang, Yanmin Qian}

\address{MoE Key Lab of Artificial Intelligence, AI Institute \\X-LANCE Lab, Department of Computer Science and Engineering
\\Shanghai Jiao Tong University, Shanghai, China  \\
{\small \tt \{zhengyang.chen, beiliu, hanbing97, zhangleying, yanminqian\}@sjtu.edu.cn}}%
\begin{document}
\maketitle

\begin{abstract}

This technical report describes the SJTU X-LANCE Lab system for the three tracks in CNSRC 2022. In this challenge, we explored the speaker embedding modeling ability of deep ResNet (Deeper r-vector). All the systems are only trained on the Cnceleb training set and we use the same systems for the three tracks in CNSRC 2022. In this challenge, our system ranks the first place in the fixed track of speaker verification task. Our best single system and fusion system achieve 0.3164 and 0.2975 minDCF respectively. Besides, we submit the result of ResNet221 to the speaker retrieval track and achieve 0.4626 mAP. More importantly, we have helped the wespeaker \cite{wang2023wespeaker} toolkit reproduce our result: 
\href{https://github.com/wenet-e2e/wespeaker/tree/master/examples/cnceleb/v2}{https://github.com/wenet-e2e/wespeaker}.

\end{abstract}

\section{Data Usage}
\label{sec:data_usage}
For all three tracks in the CNSRC 2022, we follow the same data usage setup. 797 speakers from CN-Celeb1 dev \cite{fan2020cn} and 1996 speaker from CN-Celeb2 \cite{li2022cn} are used as the training data.

\subsection{Data Processing}
As shown in \cite{li2022cn}, there are many short utterances less than 2s in CN-Celeb dataset. In our experiment, we first concatenate the short utterances from the same genre and same speaker to make them longer than 5s. It should be noted that we only do this operation on the training set. After doing this operation, the training utterance number is reduced from 632,740 to 508,228. 

\subsection{Data Augmentation}
\label{ssec:data_aug}
In our experiments, three different augmentations are applied:
\begin{itemize}
    \item Additive Noise: We use the audios from MUSAN \cite{musan2015} as the additive noise in our experiment to do data augmentation.
    \item Reverberation: The impulse response from RIR\footnote{\url{https://www.openslr.org/28/}} is used to do reverberation data augmentation.
    \item Speed Perturbation: We randomly speech up or slow down an utterance with ratio 1.1 and 0.9 to do speed perturbation. The utterance with a new speed will be considered from a new speaker \cite{yamamoto2019speaker,wang2020dku,zhao2021speakin}.
\end{itemize}
We do all the data augmentations online. For each utterance in the training process, we independently decide whether to do each data augmentation with a probability 0.6.

\subsection{Acoustic Feature Extraction and Processing}
In our experiment, we extract the 80-dimensional fbank feature for each utterance and then do the utterance-wise mean normalization along the time dimension. We did not apply voice activity detection (VAD) in our experiment.

\section{System Architecture}
\label{sec:system_architecture}

\subsection{Revisiting r-vector}
r-vector is introduced in~\cite{zeinali2019but} and is the winning system of VoxSRC 2019~\cite{chung2019voxsrc}. Although different ResNet variants~\cite{xie2019utterance, cai2018exploring,chung2018voxceleb2} are proposed for the speaker embedding learning task, they differ in various aspects such as the convolutional type (1D v.s.2D), network width, kernel size, pooling methods, etc. In our experiences, we found the r-vector \cite{zeinali2019but, wang2020data} is the most stable one considering of the performance and scalability. The 34-layer r-vector architecture used in this work is shown in Table~\ref{tab:resnet34_orig}, which is nearly the same as the original r-vector, while we use 80 dimensional Fbank instead of the 40 dimensional one. The ResNet backbone is used to transform the fbank feature into deep feature representations. Then, statistic pooling \cite{snyder2017deep} is used to map the variable-length feature sequence to fix-dimensional representation. 

\begin{table}[ht!]
\footnotesize
\centering
\caption{\textbf{Model architecture for ResNet in r-vector}}
\begin{adjustbox}{width=.45\textwidth,center}
\begin{tabular}{lcc}
\hline \hline 
Layer name & Structure & Output \\
\hline 
        Input                 & --                          & 80 $\times$ \text{Frame Num} $\times$ 1  \\
        Conv2D-1              & 3 $\times$ 3, Stride 1      & 80 $\times$ \text{Frame Num} $\times$ 32 \\
        \midrule
        ResNetBlock-1         & $\begin{bmatrix} 3 \times 3, 32  \\ 3 \times 3, 32  \end{bmatrix} \times 3$  , Stride 1& $80\times \text{Frame Num} \times32$  \\
        ResNetBlock-2         & $\begin{bmatrix} 3 \times 3, 64  \\ 3 \times 3, 64  \end{bmatrix} \times 4$, Stride $2$ & $40 \times \text{Frame Num}//2 \times 64$  \\
        ResNetBlock-3         & $\begin{bmatrix} 3 \times 3, 128 \\ 3 \times 3, 128 \end{bmatrix} \times 6$, Stride $2$ & $20 \times \text{Frame Num}//4  \times 128$ \\
        ResNetBlock-4         & $\begin{bmatrix} 3 \times 3, 256 \\ 3 \times 3, 256 \end{bmatrix} \times 3$, Stride $2$ & $10  \times \text{Frame Num}//8  \times 256$ \\
        \midrule
        StatsPooling          & --                & $20 \times 256$                 \\
        Flatten               & --                & $5120$                            \\
        \midrule
        Emb Layer                & --                & $256$  \\                 
        \hline
        \hline
    \end{tabular}
    \vspace{-3mm}
\end{adjustbox}
\label{tab:resnet34_orig}
\end{table}

\subsection{Make r-vector deeper}
It has been shown in \cite{liu22h_interspeech} that impressive improvement can be achieved via deepening the speaker embedding learner, to make the r-vector more powerful, we extend it to a deeper version. The general configuration for these ``deeper r-vector'' is shown in Table \ref{table:r_vec}.

We can change parameters $(N_1, N_2, N_3, N_4)$ in Table \ref{table:r_vec} to get deep ResNet with different layers and we list the configuration for the deep ResNet used in our experiment in Table \ref{table:resnet_configuration}. Besides, we also use the DF-ResNet in \cite{liu22h_interspeech} and make it deeper to 287 layers. 

\begin{table}[ht!]
\footnotesize
\centering
\caption{\textbf{Model architecture for deep ResNet.}}
\begin{adjustbox}{width=.45\textwidth,center}
\begin{tabular}{lcc}
\hline \hline 
Layer name & Structure & Output \\
\hline 
Input & $-$ & $80 \times \text{Frame Num} \times 1$ \\
Conv2D-1 & $3 \times 3$, Stride 1 & $80 \times \text{Frame Num} \times 32$ \\
\hline 

ResNetBlock-1 & 
$\left[\begin{array}{c}1 \times 1,32 \\ 3 \times 3,32 \\ 1 \times 1,128\end{array}\right] \times N_1$, Stride 1
& $80 \times \text{Frame Num} \times 128$ \\

ResNetBlock-2 & 
$\left[\begin{array}{c}1 \times 1,64 \\ 3 \times 3,64 \\ 1 \times 1,256\end{array}\right] \times N_2$, Stride 2
& $40 \times \text{Frame Num} // 2 \times 256$ \\

ResNetBlock-3 & 
$\left[\begin{array}{c}1 \times 1,128 \\ 3 \times 3,128 \\ 1 \times 1,512\end{array}\right] \times N_3$, Stride 2
& $20 \times \text{Frame Num} // 4 \times 512$ \\

ResNetBlock-4 & 
$\left[\begin{array}{c}1 \times 1,256 \\ 3 \times 3,256 \\ 1 \times 1,1024\end{array}\right] \times N_4$, Stride 2
& $10 \times \text{Frame Num} // 8 \times 1024$ \\

\hline 
StatisticPooling & $-$ & $20 \times 1024$ \\
Flatten & $-$ & 20480 \\
\hline Emb Layer & $-$ & 256 \\

\hline \hline
\end{tabular}
\label{table:r_vec}
\end{adjustbox}
\end{table}

\begin{table}[ht!]
\footnotesize
\centering
\caption{\textbf{Configuration for different deep ResNet.}}
\begin{adjustbox}{width=.35\textwidth,center}
\begin{tabular}{l|c}
\hline 
Deep ResNet Name & $(N_1, N_2, N_3, N_4)$ \\
\hline
ResNet152 & $(3, 8, 36, 3)$ \\ 
ResNet221 & $(6, 16, 48, 3)$ \\
ResNet293 & $(10, 20, 64, 3)$ \\
\hline
\end{tabular}
\label{table:resnet_configuration}
\end{adjustbox}
\end{table}

\section{System Training}
In our experiment, we split the training process into two stages. In the first stage, we do the general classification training. In the second stage, we do the large margin finetuning, which is first proposed in \cite{thienpondt2020idlab}. At each stage, SGD is used as the optimizer to update our models. The learning rate is exponentially decreased from an initial value to a final value. 
\subsection{Stage I}
In stage I, we use the additive angular margin (AAM) \cite{deng2019arcface,xiang2019margin} loss as the training objective. As mentioned in section \ref{ssec:data_aug}, after applying speed perturbation, the speaker number is changed from 2793 to 8379. The scale and margin in the AAM loss are set to 32 and 0.2 respectively. We trained all the systems for 165 epochs. In each epoch, we go through the whole training set and randomly sample 2s segment from each utterance to build the training batch. In this stage, the initial value and final value of the learning rate are set to 0.1 and 0.00005 respectively.

\subsection{Stage II: Large Margin Finetuning}
In stage II, we do the large margin finetuning \cite{thienpondt2020idlab} based on the model from stage I. In this stage, we abandon the speed perturbation augmentation and the speaker classification number is 2793. All the systems are trained for another 5 epochs. Besides, we randomly sampled 6s segments from each utterance to construct a training batch and change the margin in AAM to 0.5. In this stage, the initial value and final value of the learning rate are set to 0.0001 and 0.000025 respectively.\section{System Scoring}
\subsection{Scoring for Speaker Verification}
\label{ssec:scoring_ways}
In our experiment, we use cosine similarity as the scoring method. Besides, we also add the adaptive score normalization \cite{cumani2011comparison} and we set the imposter cohort size to 600. The imposter cohort is estimated from the training set by average the embeddings for each training speaker.
In the cnceleb evaluation trial, there are multiple utterances for each enrollment speaker. We tried three different ways to leverage the multiple utterances during the scoring for each trial pair.

\noindent \textbf{Utt-Concat}: In this strategy, we concatenate multiple utterances, which belong to the same enrollment speaker, to one long enrollment utterance. Then, we can directly get the score between enrollment utterance and test utterance using the scoring method mentioned above.

\noindent \textbf{Emb-Avg}: In this strategy, we first extract the speaker embedding for each utterance. Then, we average the embeddings belonging to the same enrollment speaker to get enrollment embedding. 

\noindent \textbf{Score-Avg}: In this strategy, for each utterance from the same enrollment speaker, we compute its score with the test utterance. Then, the scores are averaged to be used as the score between the enrollment speaker and the test utterance.

\subsection{Scoring for Speaker Retrieval}
\label{sec:retrieval_scoring}
In the speaker retrieval task, we score each enrollment utterance against all the utterances in the large data pool. The scoring method is the same as the method used in the speaker verification task. Then, the utterances in the data pool with top 10 scores are considered as the retrieval results.

\section{Results}
\subsection{Speaker Verification}
\subsubsection{Results comparison between different scoring strategies.}
In this section, we will compare different ways to score the speaker verification system. The corresponding results are listed in Table \ref{table:enroll_combine_compare}. The results show that applying asnorm after cosine scoring can consistently bring further improvement. Actually, we also did the score calibration \cite{thienpondt2020idlab} in our experiment. However, the score calibration can only improve the EER but degrade the minDCF and we abandon it finally. Besides, we also compare the different ways to leverage the multiple utterances belonging to one enrollment speaker and we have given a detailed description of these methods in section \ref{ssec:scoring_ways}. From the results, we find the Emb-Avg achieves the best result on the Cnceleb evaluation trial. The asnorm and Emb-Avg are applied to all the systems in the following sections.

\begin{table}[ht!]
\centering
\caption{\textbf{Results comparison between different scoring methods and different strategies to combine multiple utterances within one enrollment speaker.} The results are from the ResNet34 model after stage I training.}
\begin{adjustbox}{width=.45\textwidth,center}
\begin{tabular}{cccc}
    \toprule 
  
      Scoring Method & Enroll Comb &  minDCF (0.01) & EER (\%)
      \\
      \hline
      Cosine & Utt-Concat & 0.4391 &  7.305 \\
      Cosine & Emb-Avg & 0.4004 &  6.922 \\
      Cosine + ASnorm & Utt-Concat & 0.4035 &  7.085 \\
      Cosine + ASnorm & Emb-Avg & \textbf{0.3707} &  \textbf{6.590} \\
      Cosine + ASnorm & Score-Avg & 0.4419 &  6.759 \\
    \bottomrule
\end{tabular}
\label{table:enroll_combine_compare}
\end{adjustbox}
\end{table}

\begin{table*}[ht]
\centering
\caption{\textbf{Results for different systems.} The FNR and FPR denotes the false negative rate and false positive rate. We get the FNR and FPR by setting the score threshold to the same threshold when we get the minDCF (0.01). LM denotes the large margin finetuning.}
\begin{adjustbox}{width=.8\textwidth,center}
\begin{threeparttable}
\begin{tabular}{cccccc}
    \toprule 
  
      System & Params \# &  minDCF (0.01) & EER (\%) & FNR (\%) & FPR (\%)
      \\
      \hline
      ResNet34 $^*$         & 6.63M & 0.3958 & 7.981 & 35.29 & 0.043 \\
      ResNet34              & 6.63M & 0.3707 & 6.590 & 31.73 & 0.054 \\
      ResNet152             & 19.8M & 0.3386 & 5.762 & 29.34 & 0.045 \\
      ResNet221             & 23.8M & 0.3270 & \textbf{5.543} & 28.08 & 0.046 \\
      ResNet293             & 28.6M & \textbf{0.3202} & 5.553 & \textbf{27.92} & \textbf{0.041} \\
      DF-ResNet      & 14.8M & 0.3361 & 6.279 & 28.83 & 0.048 \\
      \hline
      ResNet34 + LM         & 6.63M & 0.3543 & 6.221 & 30.06 & 0.054 \\
      ResNet152 + LM        & 19.8M & 0.3251 & 5.452 & 28.66 & 0.039 \\
      ResNet221 + LM        & 23.8M & 0.3179 & 5.284 & 28.27 & \textbf{0.035} \\
      ResNet293 + LM        & 28.6M & \textbf{0.3164} & \textbf{5.227} & 27.82 & 0.038 \\
      DF-ResNet + LM & 14.8M & 0.3185 & 6.117 & \textbf{27.46} & 0.044 \\
      \hline
      Fusion                &  -    & 0.2975 & 4.911 & 25.28 & 0.045 \\
        \bottomrule
\end{tabular}
\begin{tablenotes}\footnotesize
\item {\footnotesize We didn't apply speed perturbation on ResNet34 $^*$ as a comparison.}
\end{tablenotes}
\end{threeparttable}
\label{table:totol_result}
\end{adjustbox}
\end{table*}

\subsubsection{Results for different systems}
Here, we listed results for all the systems in Table \ref{table:totol_result}. As we expected, the deep ResNet has a strong modeling ability and can avoid overfitting problem at the same time. For the submission of the speaker verification task, we fuse all the systems with large margin finetuning based on their performance. Besides, we also list the false negative rate (FNR) and false positive rate (FPR) in Table \ref{table:totol_result}. We get these two values by setting the score threshold to the same threshold value when we get minDCF (0.01). The results in Table \ref{table:totol_result} show that the minDCF is positively correlated with FNR and we can interpret the minDCF (0.01) as how the system performs on FNR in the case of very low FPR.

\subsection{Speaker Retrieval}
For the retrieval, we use the ResNet221+LM system in Table \ref{table:totol_result} to score each enrollment utterance against all the test utterances. Then we get the results with top 10 scores following the strategy described in section \ref{sec:retrieval_scoring}.

\bibliographystyle{IEEEbib}
\bibliography{Odyssey2022_BibEntries}

\end{document}